\documentclass[12pt]{article}
\usepackage{scicite}
\usepackage{graphicx}
\usepackage{times}
\usepackage{xcolor}

\newenvironment{sciabstract}{%
\begin{quote} \bf}
{\end{quote}}

\title{Chirality-selective optical transport of nanoparticles in the evanescent field of a nanofiber}

\author
{Georgiy Tkachenko,$^{1,2}$ Akiyoshi Suda,$^{1}$ Hyo-Yong Ahn$^{3}$, Ki Tae Nam$^{4}$,\\
Hiromi Okamoto$^{3}$, Mark Sadgrove$^{1\ast}$\\
\\
\normalsize{$^{1}$ Department of Physics, Faculty of Science, Tokyo University of Science,}\\
\normalsize{1-3 Kagurazaka, Shinjuku-ku, Tokyo 162-8601, Japan,}\\
\normalsize{$^{2}$University of Bordeaux, CNRS, Laboratoire Ondes et Matière d’Aquitaine,}\\
\normalsize{Talence F-33400, France,}\\
\normalsize{$^{3}$Institute for Molecular Science, National Institutes of Natural Sciences, }\\
\normalsize{38 Nishigonaka, Myodaiji, Okazaki, Aichi 444-8585, Japan}\\
\normalsize{$^{4}$ Department of Materials Science and Engineering, Seoul National University,}\\
\normalsize{ 1 Gwanak-ro, Gwanak-gu, Seoul 08826, Republic of Korea.}\\
\\
\normalsize{To whom correspondence should be addressed;}\\
\normalsize{$^\ast$E-mail: mark.sadgrove@rs.tus.ac.jp}
}

\date{}

\begin{document} 

\baselineskip24pt 

\maketitle 

\begin{sciabstract}
Optical nanofibers are waveguides known for their unique property to produce intense evanescent fields which have subwavelength transverse confinement easily extendable over thousands of wavelengths along the fiber axis. Moreover, circularly polarized fundamental modes of a nanofiber are chiral, that is, lacking mirror symmetry. Here, we use these two properties to demonstrate chirality-selective optical transport of a waterborne chiral material object - a chemically synthesized gold nanocube with twisted faces. Our experiments, supported by numerical simulations, show that right- and left-handed circularly polarized modes produce clearly distinct velocities of optically trapped nanocubes along the nanofiber axis, whereas non-chiral gold nanospheres of a similar size do not show any such dissymmetry. Furthermore, using a counterpropagating mode configuration, the non-chiral component of the optical force can be effectively zeroed out, yielding selective forward and backward transport of chiral nanocubes. In addition, the chiral optical force was found to be significant even for particle ensembles with natural variations in size and form, showing average behavior in agreement with numerical simulations. This is a clear implementation of optical separation of chiral enantiomers at the scale of 100 nm. Further development towards waveguide-assisted enantio-selective manipulation approaching the molecular scale can be envisaged.  
\end{sciabstract}

\section*{Introduction}
In the macroscopic world, chirality (for example, the handedness of a screw) can easily be exploited to control the translational degree of freedom of an object (the travel direction of the rotating screw, in this example). At the microscale, where  objects are typically manipulated contactlessly, directional translation of objects determined by their chirality is much more challenging. One important method is the use of chiral light~\cite{okamoto2022optical, lininger2023chirality} to manipulate chiral particles. For instance, microspheres of cholesteric liquid crystals exhibit chirality-selective propulsion~\cite{tkachenko2013spin, tkachenko2014optofluidic}, trapping~\cite{tkachenko2014helicity}, and rotation~\cite{rafayelyan2016reflective, tkachenko2017spin} on illumination with circularly polarized (CP) light. Chiral optical manipulation has also been demonstrated with micrometer long carbon nanotubes~\cite{PhysRevLett.101.127402,ajiki2009size,spesyvtseva2015chirality} - objects that are chiral (molecular helices) at the nanoscale. Recently, circular polarization dependent optical gradient forces have been demonstrated with chiral nanoparticles (CNP), namely chemically synthesized gold cubes with twisted faces~\cite{yamanishi2022optical}. While randomly moving in a water dispersion, D-forms (L-forms) of these nanocubes tend to be more strongly (weakly) localized in the focal spot of a left-handed circularly polarized (LCP) laser beam, and the picture reverses for the right-handed circular polarization (RCP).

The trend of downsizing the manipulated objects in studies of chiral optomechanics is not arbitrary - it pursues the goal of molecular-scale enantio-separation, that is sorting or selective manipulation of molecular enantiomers (molecules differing from each other only by their handedness). This is crucial for chemical processes such as drug synthesis, and thus optical enantio-separation has attracted considerable attention in recent years~\cite{guzatov2011chiral,canaguier2013mechanical,PhysRevResearch.6.023079,rukhlenko2016completely,Champi:19,ho2017enhancing,golat2024optical}.
However, because reduction in particle size also reduces the scattering and absorption cross-sections, optical manipulation becomes weaker, and is eventually overwhelmed by thermal noise. Therefore, to keep chirality-selective manipulation efficient for nanoscale objects, it is necessary to use optical fields with tighter spatial confinement. This calls for the employment of plasmonic fields which are strongly localized at the surface of a metal~\cite{Svoboda:94,urban2014optical, quidant2007optical, doi:10.1021/jz501231h, Li:22, zhang2021plasmonic} or evanescent fields near the surface of a waveguide~\cite{skelton2012evanescent,sergides2012optically,irawati2014evanescent,daly2016evanescent,doi:10.1021/acsanm.0c00274}.

Here, we report on chirality-selective optical manipulation of gold CNPs using the radiation pressure force induced by the evanescent field of an optical nanofiber (ONF). The effective one-dimensionality of the nanofiber guided modes and strong gradient force in the evanescent field have yielded a number of recent advances in particle manipulation experiments~\cite{maimaiti2015higher, tkachenko2020light, fujiwara2021optical, sadgrove2021optically, tkachenko2023evanescent}. Here, it allows the chiral component of the radiation pressure to be observed with relative ease compared to free space experiments by measuring the speed at which CNPs are transported along the fiber axis. Furthermore, by adding a counterpropagating mode in the fiber, it is possible to control the direction of chirality-selective transport using the mode polarization alone. These results demonstrate new methods which can be used to selectively control CNPs in the sub-micron regime, representing a significant contribution to the study and application of chiral optical forces.

\section*{Results}
\begin{figure}
\centering
\includegraphics[width=9 cm]{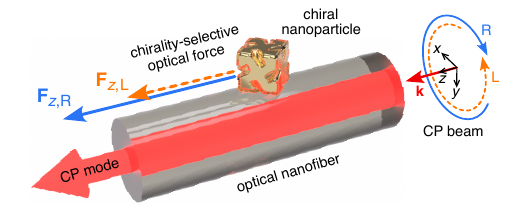}
\caption{\label{fig:concept}\textbf{Experimental concept.} A chiral gold nanoparticle is depicted at the surface of an optical nanofiber and subject to axial optical forces $F_{z,L}$ and $F_{z,R}$ due to the evanescent field of left- or right-handed circularly polarized fiber mode, respectively. }
\end{figure}
In our experiments, a single-mode ONF is submerged in an aqueous solution containing colloidal chiral nanocubes~\cite{lee2018amino}, as illustrated in Fig.~\ref{fig:concept}, where $\bf k$ is the wave vector of the input laser beam coupled to the ONF. Once in the evanescent field, such a CNP is subject to an optical force, which has three independent and potentially chiral components, $F_{x,y,z}$. The conservative component is the gradient force $F_y$ which traps the particles at the fiber surface. While its magnitude depends on the mode polarization, its sign is constant (see simulation results below) and so particles are trapped for both RCP and LCP. We note that the gradient force on such CNPs in a free-space light beam has recently been investigated~\cite{yamanishi2022optical}. The dissipative force has two components: (i)~the azimuthal force $F_x$ which produces a torque on the particle about the fiber axis, and (ii)~the radiation pressure force $F_z$ which propels particles along the fiber axis. In the present study, the transverse component (i) was found to be unobservable experimentally, in line with the theoretical predictions that state $|F_x|\ll |F_z|$ (see simulation results below). Hence, in this study we focus on the radiation pressure force which can be expressed as $F_z = F_{\rm nc} + F_{\rm c}$, where $F_{\rm nc}$ is the non-chiral part (independent of either particle or field chirality) and $F_{\rm c}$ is the chiral one. While $F_{\rm nc}$ has been well researched for metal particles (including in the setting of ONFs), studies of $F_{\rm c}$ in the nanoscale regime are still few in number.
As shown in Ref.~\cite{canaguier2013mechanical}, the chiral component of the dissipative optical force exerted on a chiral dipole is related to the chirality flow $\Phi  = (\epsilon_0 \mathcal{E}\times\dot{\mathcal{E}} + \mu_0 \mathcal{H}\times\dot{\mathcal{H}})/2$ of the field by the equation
\begin{equation}
F_{\rm c}'  = 2{\rm Im}[\chi]\Phi,
\label{eq:fchi}
\end{equation}
where $\epsilon_0$ and $\mu_0$ are the vacuum permittivity and permeability, $\mathcal{E}$ and $\mathcal{H}$ are the real parts of the electric and magnetic fields, and $\chi$ is the chiral 
polarizability in units of m$^2$s following Refs.~\cite{canaguier2013mechanical,yamanishi2022optical}. Note that in the general (non-isotropic) case $\chi$ is a tensor. 
The sign of $F_{\rm c}'$ is controlled by flipping either the handedness of the circular polarization -- positive (negative) for LCP (RCP), or the sign of Im$[\chi]$ -- negative (positive) for left- (right-) handed dipoles. Since the chirality of both the light and the particle determine the direction of the chiral force, either of these two parameters can be used to achieve a proof-of-concept demonstration of chiral optical manipulation assuming 
$F_{\rm c}\propto F_{\rm c}'$. 
Evidently, polarization of light is much easier to control compared with the geometry of synthesized CNPs. Therefore, in order to achieve the cleanest possible experimental demonstration, here we used L-form chiral particles exclusively, and switched between LCP and RCP states of the nanofiber guided light.

Following Ref.~\cite{yamanishi2022optical} where the chiral optomechanical effect was characterized by the dissymmetry of the position dispersion of CNPs in a chiral light beam, we define a dissymmetry factor for the chiral transport along the fiber axis, defined as the difference between the magnitudes of the radiation pressure optical forces for LCP and RCP divided by their mean value, that is
\begin{equation}
g_z = 2\frac{|F_{z,L}| - |F_{z,R}|}{|F_{z,L}| + |F_{z,R}|}.
\label{eq:diss}
\end{equation}
Under the dipole approximation, the chiral components have equal magnitudes and opposite signs, $F_{\rm c, R}' = -F_{\rm c, L}'$, and hence the force dissymmetry can be expressed as 
$g_z' = 2F_{\rm c,L}'/F_{\rm nc}'$,
where the sign of $g_z'$ is equal to the sign of ${\rm Im}[\chi]$ which is negative for left-handed dipoles. 
Although our CNPs are not strictly small enough for the chiral dipole approximation to hold,
it was shown in Ref.~\cite{jo2024direct} that for the particles we use, the electric and magnetic dipole moments (which contribute to the chiral dipole moment) dominate over the next highest order moment (quadrupole moment) near the chiral resonance. Thus, we expect 
$g_z\propto g_z'<0$ 
for our L-form CNPs in the spectral region near their strongest chiral response.

\subsection*{Particle analysis and modeling}
\begin{figure}
\centering
\includegraphics[width=\linewidth]{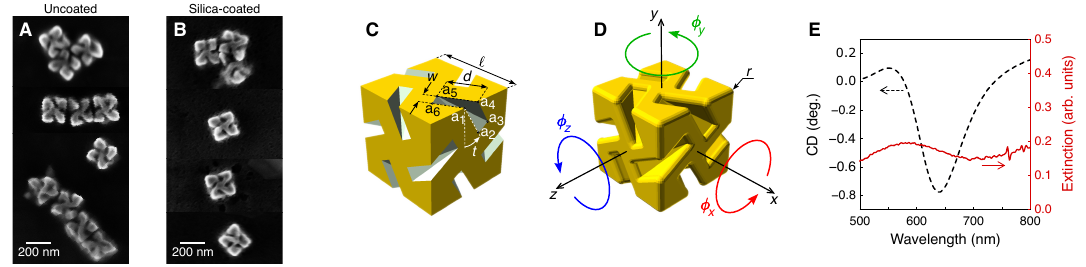}
\caption{\label{fig:exp_CD_SEM}\textbf{Chiral nanoparticles.} Same scale SEM images of pristine~(\textbf{A}) and silica-coated (\textbf{B}) CNPs. \textbf{C,~D}~Geometric model of a chiral cube, without and with rounding respectively. \textbf{E}~Measured circular dichroism and extinction spectra of coated CNPs.}
\end{figure}
The CNPs used in our experiment are shown in a scanning electron microscope (SEM) image  in Fig.~\ref{fig:exp_CD_SEM}\textbf{A}. The particles had L-form chirality (i.e. anti-clockwise twisting from the center). To reduce incidents of adhesion to the optical nanofiber, the particles were silica coated (coating thickness about 5~nm). A scanning electron microscope (SEM) image of the coated particles is shown in Fig.~\ref{fig:exp_CD_SEM}\textbf{B}. As an ensemble, the particles all tend to exhibit chirality, but also show a range of minor structural differences, particularly after coating. 

In order to numerically simulate the optical force on these particles, we used a parameterized particle model based on a modified version of the model of Lee {\it et al.}~\cite{lee2018amino}, as shown in  Fig.~\ref{fig:exp_CD_SEM}\textbf{C}. To form the model, tilted cuts were made on all twelve edges of a cube. The model then has four variable parameters: $\ell$ -- side length width, $w$ -- cut width, $d$ -- cut depth, and $t$ -- the tilt angle of the cut as measured from the normal to the edge. The sign of the chirality depends on the tilt, that is L-form corresponds to $(0<t<90^{\circ})$ and D-form to $(-90^{\circ}<t<0)$. 

As shown in Fig.~\ref{fig:exp_CD_SEM}\textbf{D}, we also rounded all edges and corners to mimic the slightly rounded nature of real CNPs. A silica coating of thickness $s$ was produced by creating a chiral cubic shell with length $(\ell+2s)$, cut width $(w-2s)$ and the same $d$ and $t$ parameters as the gold CNP. 

The particle chirality is experimentally ascertained by measuring the circular dichroism (CD) of particles in solution. Measurements of the CD along with the particle extinction spectrum are shown in Fig.~\ref{fig:exp_CD_SEM}\textbf{E}. The CD is seen to exhibit an inverted peak with a minimum between 640 and 650 nm. Optical manipulation at wavelengths near to this value is expected to produce the most vivid demonstration of chiral aspects of the optical force on the CNPs.

Due to the number of parameters in the particle model, and the memory intensive nature of the simulations, a systematic exploration of the parameter space is impractical. Instead, we made measurements of numerous particles, both uncoated and coated, in a scanning electron microscope (SEM), giving rise to the parameters shown in Table~\ref{tab:particle_params}.
Simulations of the CD and $g_z$ were first performed for a particle model using the mean parameter values. We also performed simulations where each parameter in turn was set to its mean value $\pm 1\sigma$, where $\sigma$ is the measured standard deviation of the parameter in question. This allowed us to determine the expected range of experimental measurements.

\begin{table}
\centering
\begin{tabular}{|l|l|l|l|l|l|}
\hline
 {\bf Particle type} & {\bf Total particle number} & $\ell$ (nm) & $d$ (nm)& $w$ (nm)& $t$ (degrees) \\ \hline
 {\bf Uncoated} & 42 & $180\pm10$ & $65\pm5$ & $35\pm5$ & $30\pm5$ \\ \hline
 {\bf Silica coated} & 24 & $190\pm10$ & $70\pm5$ & $35\pm5$ & $30\pm15$ \\ \hline
\end{tabular}
\caption{\label{tab:particle_params} \textbf{Summary of SEM analysis of particle parameters.} Mean values and standard deviations are given rounded to the nearest 5 nm.}
\end{table}

\subsection*{Simulation results}
\begin{figure}
\centering
\includegraphics[width=\linewidth]{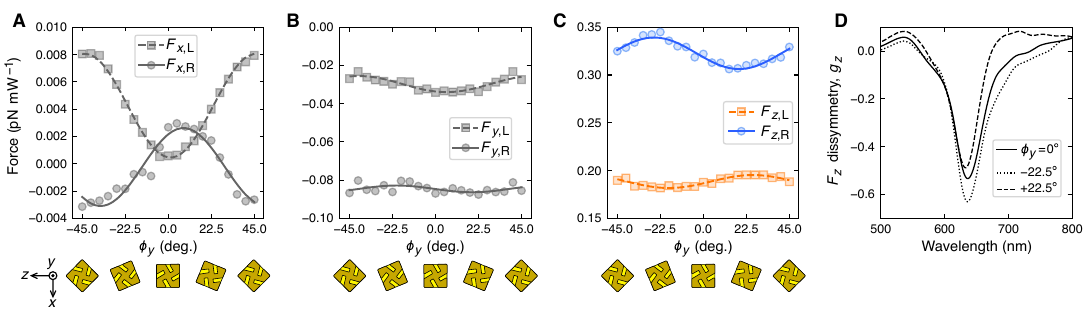}
\caption{\label{fig:sim_forces}\textbf{Simulated optical forces on an uncoated CNP near the surface of an ONF guiding a CP mode.} Panels \textbf{A}, \textbf{B}, and \textbf{C} show the azimuthal ($x$-directed), radial ($y$-directed) and axial ($z$-directed) optical force, respectively, for RCP (circles) and LCP (squares), vs. the orientation angle $\phi_y$ of the CNP. Solid and dashed curves are the best sinusoidal fits to the simulation data for RCP and LCP respectively. The sketches below indicate the CNP orientations at the corresponding $\phi_y$ values. The wavelength of the mode was 640 nm in each case, corresponding to maximal dissymmetry in $F_z$. \textbf{D}~Axial force dissymmetry, $g_z$ (Eq.~\ref{eq:diss}), calculated for the zero rotation ($\phi_y=0$, solid curve) and for the two orientations with the extrema of $F_{z}$; $\phi_y=+22.5^{\circ}$ (dashed curve) and $\phi_y=-22.5^{\circ}$ (dotted curve).}
\end{figure}
Because of the relatively large size of the particles considered here, treatments of the chiral force developed in the dipole approximation~\cite{moffitt1956optical, power1974circular,tang2010optical,govorov2010theory, genet2022chiral, golat2024optical} are not strictly applicable, and in general, we compare our experimental results with numerically calculated forces. We simulated the optical force on the CNPs using the finite difference time domain (FDTD) method of electromagnetic field propagation. The force itself was calculated using the Maxwell stress tensor applied to the calculated fields. Details of the simulations are given in the Materials and Methods Section.

In the experiments, we assume that the CNPs are in the maximally stable configuration where one face of the cube touches the ONF surface, that is, we assume $\phi_x=\phi_z=0$, see Fig.\ref{fig:exp_CD_SEM}\textbf{D}. Any deviation from this configuration should be corrected by the gradient force, $F_y$. However, we have no knowledge about the rotation of the cubes about the $y$-axis through the cube center and normal to the cube face. We therefore simulated the optical force over the whole range $-45^{\rm o}\leq\phi_y\leq45^{\rm o}$ to investigate the behavior as a function of rotation.

Figures~\ref{fig:sim_forces}\textbf{A},\textbf{B} and \textbf{C} show the simulated $x$, $y$, and $z$ components respectively of the optical force for both LCP (squares) and RCP (circles) states of the fiber mode. The curves in each of these panels are sinusoidal fits to the data. As expected, all the force components for LCP and RCP are clearly different in both mean value and their dependence on $\phi_y$. The most striking feature of the results is that the peak absolute value of the $z$-directed force is about five times larger than that for the $y$ directed gradient force, and up to two orders of magnitude larger than that of the $x$ component. This fact justifies our approach here of focusing on the $F_z$ force directed along the fiber axis and ignoring the transverse components for the present experiment. 
We also note that the variation of $F_z$ with $\phi_y$ is of order $10\%$, with its average found at $\phi_y=0$ for both LCP and RCP. This allows us to restrict the subsequent simulations to $\phi_y=0$, being confident that the results for this value reflect the average behavior.
The fitted sinusoidal curves serve as a guide for the eye.

We note in passing that among all the force components, the behavior of $F_x$ (Fig.~\ref{fig:sim_forces}\textbf{A}) shows the largest variation as a function of $\phi_y$, with a change in sign predicted in the case of RCP. Although these numerical predictions are intriguing, the small magnitude of the force in this case means they could not be confidently detected in our experiments.

In Fig.~\ref{fig:sim_forces}\textbf{D}, finite-difference time-domain (FDTD) simulations of the force dissymmetry along the $z$ axis for three values of $\phi_y$ are shown. We see that the qualitative behavior is the same for all values of $\phi_y$ and that the result for $\phi_y=0$ broadly corresponds to the average result as expected. It is also worth noting the qualitative agreement between the simulated and analytically predicted results for the force dissymmetry near the resonance, that is $g_z<0$ in the resonant dip centered around the wavelength of 660~nm.

\subsection*{Experimental results: Particle transport}
\begin{figure}
\centering
\includegraphics[width=\linewidth]{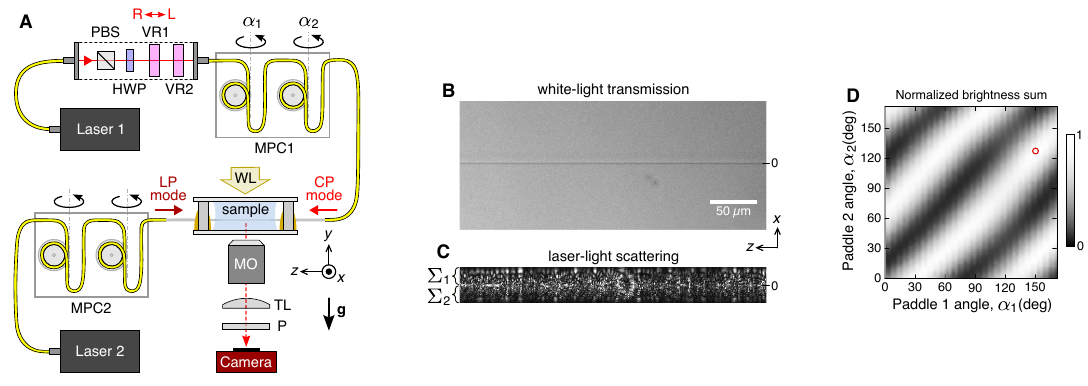}
\caption{\label{fig:setup}\textbf{Experimental details.} \textbf{A}~Schematic diagram of the optical setup (not to scale). Here PBS stands for a polarizing beam splitter; HWP -- half-wave plate; VR -- variable retarder; MPC -- motorized polarization controller, WL -- white light source in K\"{o}hler configuraiton; MO -- microscope objective; TL -- tube lens; P -- linear polarizer; LP -- linear polarization; CP -- circular polarization; $\bf g$ -- gravitational acceleration.  \textbf{B}~Camera image of the ONF waist region in transmission under WL illumination. \textbf{C}~Image of the laser light scattered from the same part of ONF placed slightly out of focus (closer to MO). $\Sigma_1$ and $\Sigma_2$ are the brightness sums in the top and bottom halves of the image. \textbf{D}~Typical map of the total brightness of the imaged ONF-scattered laser light versus the orientation angles of the MPC paddles. The red circle indicates the position of maximum brightness sum.}
\end{figure}
The numerically predicted maximum force dissymmetry of $g_z\approx-0.5$ corresponds to $40\%$ difference between the optical forces and the corresponding velocities (see below) for right- and left-handed CP modes. From previous experiments~\cite{sadgrove2021optically} on optical transport of gold nanoparticles along single-mode ONFs, we expect velocities of order $200\,\mu$m~s$^{-1}$ for the typical mode power of 10~mW in the visible range. The corresponding expected velocity difference of about $80\,\mu$m~s$^{-1}$ can be easily detected via an optical microscope imaging the laser light scattered by the transported particle. We designed an experiment to make such measurements as depicted in Fig.~\ref{fig:setup}\textbf{A} and described in detail in the Materials and Methods Section.

Of principle importance is our ability to accurately control the polarization of light in the fiber. In particular, for directional transport experiments, a single CP mode was injected in the $+z$ direction, while for oscillating transport experiments, an additional linearly polarized (LP) mode was injected in the $-z$ direction. In order to ensure accurate control of the mode polarization in the ONF waist region (shown by the transmission white-light image and the scattered laser-light image in Fig.~\ref{fig:setup}\textbf{B} and \textbf{C}), we applied the two-step compensation method~\cite{lei2019complete,tkachenko2019polarization} where the first step (mapping of the horizontal polarization) in this setup is realized by employing motorized polarization controllers (MPC), see Materials and Methods for more details.

\begin{figure}
\centering
\includegraphics[width=\linewidth]{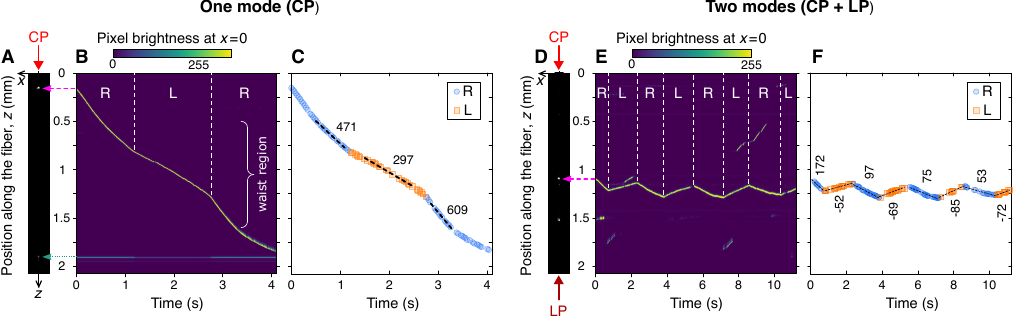}
\caption{\label{fig:CP_modulation}\textbf{Polarization-dependent dynamics of a CNP.} \textbf{A-C}~Transport by a single CP input mode, switching between right- and left-handed polarization states. (\textbf{A})~First frame of the video recording where the bright spot indicated by the dashed arrow represents the transported CNP. The dotted arrow below indicates a fixed smaller scatterer. \textbf{B}~Sequential compilation of lines extracted from the video frames at $x=0$ (ONF axis). Here, R indicates RCP and L indicates LCP. \textbf{C}~Measured CNP position vs. time. The dashed lines are the best linear fits to the data in the corresponding uniform pieces of the track. \textbf{D-F}~Chirality-selective directional transport of a CNP by two counterpropagating modes, one of which is fixed at LP, while the other alternates between RCP and LCP states. In both one-mode and two-mode configurations, the CP mode wavelength is 660~nm and the transmitted power is fixed at 6~mW. The LP mode has 637~nm wavelength and approximately the same constant power fine-tuned to achieve near-balanced particle dynamics prior to measurement. The numbers next to the lines in \textbf{C,~F} indicate the fitted speed in the units of $\mu$m~s$^{-1}$.}
\end{figure}

As demonstrated in Figure~\ref{fig:CP_modulation}, when the polarization of the CP mode is switched between R and L states (by means of the variable retarder, VR1, shown in Fig.~\ref{fig:setup}\textbf{A}), the transport velocity of a CNP is visibly modulated. Indeed, the slope of the track, which indicates the particle position vs. time, decreases from $471\,\mu$m~s$^{-1}$ for the RCP state down to $297\,\mu$m~s$^{-1}$ for LCP, and then rebounds to $609\,\mu$m~s$^{-1}$ once the polarization is switched back to RCP, see Figure~\ref{fig:CP_modulation}\textbf{B}. These values were extracted by performing linear fits to the data (dashed lines in Fig.~\ref{fig:CP_modulation}\textbf{C}). Note that we limited this analysis to the waist region where the ONF diameter was nearly constant. For the fibers used here, the taper design gives a nominal 1-mm-long region of diameter 400~nm with about 10$\%$ variation between samples, as verified by SEM analysis of freshly prepared ONFs. Outside this region, the fiber diameter increases exponentially leading to a fall off in the evanescent field and thus a reduction of the optical force leading to slowing of the particle, as seen near the start and finish of the track in Fig.~\ref{fig:CP_modulation}\textbf{B,~C}.

The measured steady state velocities $v$ of the transported particles allowed us to characterize the chiral optical forces. Indeed, in the present case the Reynolds number is low, thus the flow around the particle is laminar and the force is proportional to $\eta v$, where $\eta$ is the dynamic viscosity of water. Assuming that $\eta$ is independent of the polarization state (that is, temperature difference between the states is negligible), we express the force dissymmetry (Eq.~\ref{eq:diss}) as
\begin{equation}
g_z=2\frac{v_L-v_R}{v_L+v_R}\,,
\end{equation}
where $v_L$ and $v_R$ are respectively the LCP and RCP steady state velocities determined from the best linear fit to the particle track, and the modulus operators are omitted because velocities (and the forces, as follows from $F_z$ simulations) have the same sign. Taking the average of the velocities in the two RCP regions in Fig.~\ref{fig:CP_modulation}\textbf{B,~C} as $v_R$, we find $f = -0.58$ for this track. Notably, this value is close to the numerically predicted one, see Fig.~\ref{fig:sim_forces}\textbf{D}.

The above demonstration proves that it is possible to manipulate the transport of chiral nanocubes contingent on their chiral property. However, the change in velocity measured due to the chiral optical force was still small compared to the overall transport velocity. In order to perform more useful manipulation,  it is desirable to isolate the chiral optical force by canceling the non-chiral component. This can be done by introducing a constant counterpropagating mode (here, with $x$-oriented linear polarization) into the ONF, as depicted in Fig.~\ref{fig:CP_modulation}\textbf{D}. (Note that the wavelength of the two modes is chosen to be different for experimental convenience, as it allows the modes to be separated at the fiber ends by filtering). By varying the power in this counterpropagating mode, the transport velocity for RCP and LCP may be adjusted, and for the appropriate power setting the non-chiral part of the force, $F_{\rm nc}$, can be completely canceled. In this regime, the direction of transport is different for each polarization handedness. It is therefore possible to keep the particle position at the ONF waist oscillating by altering the polarization periodically. We show a realization of this effect in Fig.~\ref{fig:CP_modulation}\textbf{E,~F} where the CNP position is seen to oscillate within approximately 0.25~mm. This is effectively a proof-of-concept demonstration of chiral optical sorting of CNPs along a nanofiber waveguide.

\subsection*{Variation of experimental parameters}
\begin{figure}
\centering
\includegraphics[width=\linewidth]{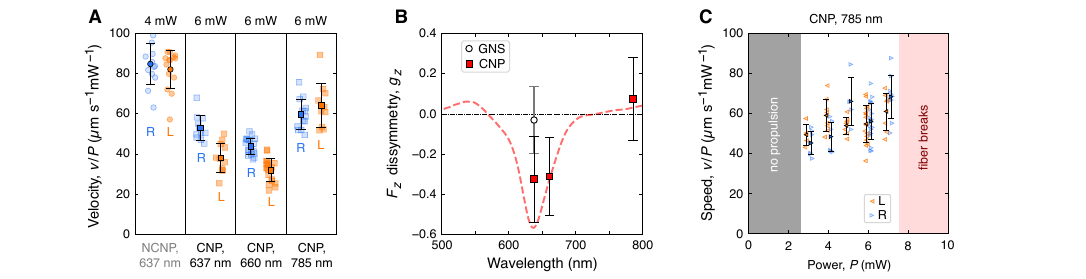}
\caption{\label{fig:velocity}\textbf{Experimental results for multiple chiral and nonchiral nanoparticles transported by a single CP mode.} In \textbf{A}, leftmost panel, measured transport velocities for NCNPs (gold nanospheres) are shown for RCP (blue circles, label R) and LCP (orange circles, label L) states. The subsequent panels from left to right  show the results for CNPs at mode wavelengths of 637~nm, 660~nm, and 785~nm respectively. In each panel, blue (orange) squares show results for RCP (LCP), and black squares with error bars show the mean and $\pm1\sigma$ of the data sets, respectively. \textbf{B}~Axial force dissymmetry for each data set, with the while circle showing the result for NCNPs, and the red squares those for CNPs. The overlaid dashed curve shows the simulation result for the geometric parameters’ values listed in Table~1. \textbf{C}~Power dependence of the variance in CNP velocity measurements at the mode wavelength of 785~nm. The shaded gray region shows the regime where the power is too low to produce particle transport, while in the pink shaded region the fiber was found to break immediately or shortly after the first CNP was captured.}
\end{figure}
For the chiral optical force to be useful in controlling optical transport in realistic situations, $F_{\rm c}$ should dominate over fluctuations in $F_z$ due to other experimental factors. Principle among these are the inhomogeneity of particle size and form. The effect of small structural deviations between individual CNPs on the chiro-optical responses has been discussed previously~\cite{cho2020uniform,spreyer2022second}. These factors lead to differences in both $F_{\rm nc}$ and $F_{\rm c}$ from particle to particle.

To assess the relative size of such effects compared to the chiral force effect, we performed measurements of particle velocity for at least 10 individual nanoparticles for both LCP and RCP modes (in the one-mode regime). We also checked the effect of wavelength and performed control measurements using non-chiral nanoparticles (NCNPs) -- gold nanospheres of 150 nm diameter -- at 637 nm which gives the largest chiral response for our CNPs. The results for these control measurements are shown in the left-most panel of Fig.~\ref{fig:velocity}\textbf{A}. Although a wide spread is seen for the measured velocities in the case of both LCP and RCP, there is no significant difference seen in the mean velocities for the two polarizations, as may be seen by the almost total overlap of the error bars. This null experiment with NCNPs was essential as it verified the validity of $F_{\rm nc,\,L}=F_{\rm nc,\,R}$ assumption in our setup.  

Conversely, in the results for CNPs at both 637 nm and 660 nm, the mean value of the measured velocities lies outside the $\pm1\sigma$ interval, with the RCP mode on average producing larger velocities than LCP for both wavelengths. Furthermore, in agreement with the numerical simulations, when the mode wavelength was set to 785 nm, the difference in velocities between data for LCP and RCP was much smaller compared to the spectral dip (637 nm), and was, in fact, insignificant compared to the standard deviation of the data.

The results for 637 nm and 660 nm show that even in the presence of particle geometry variation which affects the optical force, the chiral force still dominates. To confirm this, we also calculated the force dissymmetry from the above data, see Fig.~\ref{fig:velocity}\textbf{B}. As expected, the dissymmetry in the case of NCNPs is zero within the error of the measurement. Simulation results for the mean geometric parameters listed in Table~1 are overlaid as a red dashed line. Although the experimentally measured dissymmetries have a large variance, the mean $g_z$ values are seen to be close to their numerically predicted ones. 
We also note that the wavelength dependence of the force dissymmetry broadly follows the dependence shown by the CD in Fig.~\ref{fig:exp_CD_SEM}\textbf{E}, which finding suggests that the key physical mechanism behind the force dissymmetry is the chirality-selective absorption of light by CNPs.

Finally, we explore how the measured velocities depend on the power of the fiber mode, $P$, see 
Fig.~\ref{fig:velocity}\textbf{C}. The data were collected at 785~nm, which wavelength corresponds to the 
largest transport velocities. For power values in the pink region of Fig.~\ref{fig:velocity}\textbf{C}, fiber damage occurred limiting the maximum power which could be used in experiments. Notably, the normalized velocity, $v/P$, is not constant with power but slowly increases, thus indicating a nonlinear process. The most likely explanation for this behavior is increased heating of the solution and therefore reduction of $\eta$ (besides possible thermotaxis which is beyond the scope of this study).

The observation of fiber damage allows us to infer a rough probable value for the temperature of the water surrounding CNPs near to the fiber. We first note that such ONFs immersed in pure water can reliably sustain 50-fold higher optical powers of the guided mode~\cite{tkachenko2023evanescent}, hence the observed breaking at the modest power levels of $P>7$~mW must be due to the presence of trapped CNPs. Let us assume that the fiber breaks due to cavitation~\cite{arita2013cavitation} when the local temperature near to the fiber surface, induced by heating of the CNPs, rises to 100$^{\rm o}$C, and the ensuing phase transition leads to a sudden change in surface tension. At 7~mW, the particle velocity is about $70$ $\mu$m s$^{-1}$ mW$^{-1}$ (see Fig.~\ref{fig:velocity}\textbf{C}). The dynamic viscosity of water at 100$^{\rm o}$C is $\eta=0.2814$~mPa~s. On the other hand, at 3~mW the velocity is about $45$ $\mu$m s$^{-1}$ mW$^{-1}$. Then, assuming  laminar flow, so that the optical force is proportional to $\eta v$, the power normalized forces may be equated allowing us to derive $\eta$ at 3~mW as $\eta=0.2814\times70/45=0.4377$~mPa~s. This value of $\eta$ corresponds to a water temperature near to the fiber of 64$^{\rm o}$C.

\section*{Discussion}
The results presented above show that it is possible to manipulate a particle based on its chiral property using the evanescent portion of chiral light propagating in a nanofiber. The effective one dimensional nature of the system reduces the degrees of freedom of particle movement so that motion along the fiber axis can be easily measured. Indeed, this method offers sufficient sensitivity that the difference in chiral optical force between left and right handed circular polarizations can be distinguished in the one-mode particle transport observations alone. Measurements of the effect at different wavelengths agreed quantitatively with simulation results, and the same experiment run using non-chiral particles produced no force dissymmetry, as expected. 

Moreover, a counterpropagating mode configuration enabled isolation of the chiral optical force, allowing us to tie the direction of chiral nanoparticle transport to the handedness of the chiral light in the nanofiber. This method holds promise for more sophisticated manipulation e.g. chirality-selective nanoparticle trapping and sorting, which we aim to explore in the future. 

Another important prospect is to further reduce the size of the nanoparticles which can be manipulated by these techniques, with the ultimate goal being the selective optical manipulation of chiral molecules. Given the current signal to noise ratio, it should be possible to measure velocities down to 10 $\mu$m~s${^{-1}}$ before the thermal noise swamps the velocity signal. This corresponds to forces of $\sim 15$ fN and to a reduction in particle volume by $\sim10$ times. The corresponding reduction in particle size is approximately 2 times, reaching the sub 100 nm size regime. Further downsizing requires more optical power, which can become problematic when using metallic nanoparticles near the plasmonic resonance, but should not be an issue for dielectric enantiomers such as drug molecules. Importantly, the relatively simple method and the promising nature of the results shown here suggest that the nanofiber platform is ripe for further applications to chirality-selective optical manipulation.

\section*{Acknowledgments}
This work was supported by KAKENHI Grant-in-Aid for Transformative Research Areas (Grant No. JP22H05135) and for Scientific Research (A) (Grant No. JP21H04641).

\section*{Materials and methods}
\subsection*{Simulations}
Numerical simulations were performed using a commercial implementation of the finite-difference time-domain method (Lumerical FDTD), with force calculations made from numerical evaluation and integration of the Maxwell stress tensor 
\[
T_{ij}={\frac {1}{4\pi }}\left(E_{i}E_{j}+H_{i}H_{j}-{\frac {1}{2}}\left(|\mathbf{E}|^{2}+|\mathbf{H}|^{2}\right)\delta _{ij}\right)
\] 
where $E$ and $H$ are amplitudes of the electric and magnetic fields $\mathbf{E}$ and $\mathbf{H}$ respectively, and the indices $i$ and $j$ range over the $\{x,y,z\}$ components.

We simulated the chiral cube nanoparticle using a parametric model in OpenSCAD software following the same approach as in previous studies of such CNPs~\cite{lee2018amino,yamanishi2022optical,cho2020uniform}. The refractive index of CNP was set using a model of gold based on measurements published in the CRC Handbook of Chemistry and Physics~\cite{haynes2016crc}. 

The nanofiber was modeled as a constant diameter silica cylinder centered at $(x=0,y=0)$ and parallel to the $z$ axis. The simulation included a $10\,\mu$m length of the fiber extending through the perfectly matched layer simulation boundaries. Field evaluations for the calculation of the stress tensor take place at the faces of a box which surrounds the CNP, but does not intersect with the nanofiber. The simulation meshing was non-uniform with a minimum grid size of 5~nm inside the CNP. In addition, we applied the conformal mesh refinement method~\cite{yu2001conformal}.

The particle was typically placed between 0 and 20 nm above the fiber surface in the simulations. Although the exact distance above the surface does affect the absolute value of the force, our numerical tests showed that the value of the force dissymmetry $g_z$ was not significantly affected by this placement, due to the chirality of the evanescent field being essentially independent of position.

LCP and RCP fiber modes were created by overlapping $x$- and $y$-polarized fundamental modes of the fiber with a phase difference of $\pi/2$ or $-\pi/2$ respectively. The modes propagated in the $+z$ direction. 

Scattering and absorption cross-sections of the CNPs were calculated using the supplied and purpose-made total-field scattered-field (TFSF) source in the commercial software.

\subsection*{Nanoparticle fabrication}
The chiral nanoparticles used in this study were prepared by peptide-directed synthesis method~\cite{lee2018amino}. Pre-synthesized octahedral seeds were dispersed in an aqueous solution of hexadecyltrimethylammonium bromide (CTAB, 1 mM) before use. The growth solution was prepared by adding CTAB solution (0.8 mL, 100 mM) and aqueous gold chloride trihydrate solution (0.1 mL, 10 mM) to deionized water (3.95 mL). Aqueous ascorbic acid solution (0.475 mL, 100 mM) was then added to reduce Au$^{3+}$ to Au$^{+}$. The growth of chiral nanoparticles was initiated by adding aqueous L-glutathione solution (0.005 mL, 5 mM), followed by the addition of octahedral seed solution (0.05 mL). After stirring for 60 s, the growth solution was kept undisturbed at 30 °C for 2 h. The solution was centrifuged twice and redispersed in CTAB solution (1 mM). 

For the preparation of silica-coated nanoparticles, an aqueous solution of methoxy polyethylene glycol thiol (mPEG-SH, 5 kDa, 0.03 mL, 0.25 mM) was added to the as-synthesized chiral nanoparticle solution (1 mL) and stirred continuously for 30 min. The mPEG-modified nanoparticle solution was centrifuged three times and redispersed in ethanol (0.58 mL). Deionized water (0.167 mL) and ethanol solution of ammonia (0.07 mL, 2 M) were added while stirring continuously. The reaction was started by adding isopropanol solution of tetraethyl orthosilicate (0.014 mL, 10 vol$\%$) and the solution was stirred continuously for 2 h. The silica-coated nanoparticle solution was centrifuged three times with ethanol. For the transport experiment and further characterization, the nanoparticles were redispersed with deionized water. The circular dichroism spectrum of the silica-coated nanoparticle solution was measured using a commercial spectrophotometer (J-1500, JASCO corp.) with a 1 mm path length quartz cell.

Non-chiral particles were gold nanospheres with the nominal diameter of 150~nm, obtained commercially from Nanopartz (A11-150-BARE-DIH).

\subsection*{Nanofiber fabrication and mounting}
Optical nanofibers were manufactured in-house by tapering a commercial optical fiber (780HP by Thorlabs, Inc.) using a standard heat-and-pull technique~\cite{birks1992shape,ward2006heat} with hydrogen flame brushing. The waist region of the ONF had a typical diameter of 400~nm with a variation of about 10$\%$ between pulls. The adiabaticity of the pulling process was verified by measuring the transmission of the coupled laser light at 780~nm wavelength.

Once fabricated, the fiber was immersed in a drop (around 0.1~mL) of Milli-Q water and fixed to the a microscope glass slide using an ultraviolet light-cured adhesive. In order to improve bright-field imaging of the fiber and to eliminate the adverse effects of the liquid evaporation, the immersed part of the fiber was covered with a second glass slide (150-nm-thick cover slip) being in contact with water and supported at the corners by plastic patches around 1.5~mm thick. After being thus prepared, the sample was mounted on the 3D micrometric stage of the optical setup and connected to the fiber cables by fusion splicing. We performed bright-field transmission imaging of the ONF in order to locate its waist region and realized the polarization compensation protocol for both forward ($+z$) and backward ($-z$) directions of propagation. Once these preparatory steps were done, we carefully lifted the cover slide and deposited a small drop (around 10~$\mu$L) of the aqueous solution with CNPs (or NCNPs) on top of the ONF, and repositioned the slide in order to stop diffusive flows and to protect the sample from dust.

\subsection*{Optical setup}
The experimental setup is sketched in Fig.~\ref{fig:setup}\textbf{A}. As the laser sources, we used fiber-coupled laser diodes LP637-SF60 ($P\leq70$~mW, wavelength 637~nm), LP600-SF60 ($\leq60$~mW, 660~nm), and LP785-SF100 ($\leq100$~mW, 785~nm), all from Thorlabs, Inc. For better mechanical stability and smaller footprint, the laser-linked parts of the setup were designed to be all-fiber, except the short free-space segment in the fiber bench (FB-76 with PAF2-2B fiber ports by Thorlabs, Inc.) containing pin-mounted optical elements for the polarization control at the CP-mode side. The fiber-coupled diode Laser~1 of the chosen wavelength was connected directly to the fiber bench whose output port was connected to a single-mode patch cable (P1-630Y-FC-1 by Thorlabs, Inc.) fixed on the paddles of MPC1. The cleaved end of this patch cable was fusion-spliced to one of the fiber pigtails of the ONF sample, and all lose fibers were taped to the optical table to ensure the stability of polarization and power of the fiber modes. In the two-mode scheme (Fig.~\ref{fig:sim_forces}\textbf{D}), the second fiber pigtail of the sample was spliced to the patch cable running through MPC2 and connected (through the second fiber bench, not shown) to the fiber-coupled output of the Laser~2, the source of the counterpropagating LP mode. The mode power was measured in transmission through the ONF (S130C photodiode sensor with the fiber bench mount FBSM, by Thorlabs, Inc.), while no particles were transported.

In order to locate the waist region, the water-immersed ONF sample was imaged in transmission using a microscope objective lens (Nikon Plan Fluor 40x/0.60) under condensed illumination from a light-emitting diode. Light collected by the the objective lens was focused by a plano-convex tube lens (focal length 150~mm) and imaged by a CMOS camera (Zelux CS165CU by Thorlabs, Inc.) fixed on a rotating mount which was adjusted in order to have the fiber parallel to the horizontal sides of the image.

\subsection*{Polarization control}
Two motorized polarization controllers, MPC1 and MPC2, and a liquid crystal variable retarder, VR2 (LCC1513-B by Thorlabs, Inc., on a custom 3D-printed pin-mount), were the only adjustable elements in the polarization compensation stage of the experiment. The key to this procedure was imaging of the laser light scattered by natural imperfections of the ONF waist in pure water.

Following the two-step method~\cite{tkachenko2019polarization}, we first found and set the optimum angles $\alpha_1$ and $\alpha_2$ (marked by the red circle in Fig.~\ref{fig:setup}\textbf{D}) for the MPC1 paddles which were wrapped in three loops of the fiber patch cable (to ensure the complete coverage of the Poincar{\'e} sphere). At these angle values, the total polarization transformation in the fiber was such that the horizontal state (provided by the polarization beam splitter, FBT-PBS052, by Thorlabs, Inc.) was transferred from the fiber bench to the ONF waist unchanged. The horizontal (along the $x$ axis in the sample frame, see Fig.~\ref{fig:setup}\textbf{A-C}) polarization was verified by locating the maximum of the brightness sum of the laser-scattering image (Fig.~\ref{fig:setup}\textbf{C}). In order to maximize the contrast of the brightness sum map, the longitudinal component of the light field was blocked by a linear polarizer, P. We note that compared to the previously reported realization of this first compensation step by random manual rotation of two quarter-wave plates~\cite{lei2019complete,tkachenko2019polarization}, the automated approach realized here is quicker and much less sensitive to the operator's errors, and therefore it is much more accurate.

For the second step of the compensation procedure, we inserted a half-wave plate, HWP, and the variable retarder, VR2. The HWP was pre-aligned using a free-space polarization analyzer (PAX1000VIS, by Thorlabs, Inc.) to produce the diagonal linear polarization after the PBS. Then the retardance of VR2 was tuned such that the fiber mode at the ONF waist was also diagonally polarized (that is, having the electric field in the longitudinal plane tilted at $+45^{\circ}$ to $x$ axis). This was checked by locating the absolute minimum of $\Delta = \Sigma_1-\Sigma_2$, where $\Sigma_1$ and $\Sigma_2$ are the brightness sums in the top ($x>0$) and bottom ($x<0$) parts of the scattering image (Fig.~\ref{fig:setup}\textbf{C}), respectively. Once these two steps were done, any polarization state -- including CP states -- was considered to be transferred from the fiber bench to the ONF waist unchanged~\cite{lei2019complete}. We than inserted (directly after the HWP) another full-wave variable retarder, VR1 (same as VR2), operated by a custom controller (square-wave generator based on OP07 operational amplifier, oscillation frequency 2~kHz) with two preset voltage values corresponding to RCP and LCP states (verified by PAX1000VIS in free space). During the experiments, the CP mode polarization could be toggle-switched between these two states.

For the two-mode scheme, only fixed horizontal LP state was required. It was achieved by the partial compensation procedure where we located and set the paddle angles of MPC2 corresponding to the maximum brightness sum of the scattering image with only the Laser~2 being coupled to the ONF. The MPCs were operated by a custom executable program (developed in C$\#$ language for Windows OS, requires the Kinesis software package by Thorlabs, Inc.). The acquisition of the scattering image is realized by capturing the portion of the computer screen where the (appropriately adjusted) camera image is displayed.
 
\subsection*{Measurement and analysis of particle transport}
Scattered laser light from nanoparticles trapped and transported in the ONF evanescent field was recorded by the CMOS camera to movie files which were the raw data from the experiment. In order to maximize the field of view and the frame rate, we used the lower-magnification objective lens (Olympus LU Plan Fluor 5x/0.15) and cropped the image to fit the ONF and about $\pm50\,\mu$m in the $x$ direction. The position of the particle along the fiber axis in each video frame was extracted from the raw data by a custom Python code. The best linear fit to these positions vs. time, as shown in Fig.~\ref{fig:CP_modulation}\textbf{C,~F}, gave the transport velocity as the slope of the line.

The $\pm1\sigma$ confidence ranges for $v_L$ and $v_R$ shown in  Fig.~\ref{fig:velocity}\textbf{A,~C} are respectively $\sigma_L$ and $\sigma_R$, calculated as the standard deviations of the measured velocities from their mean values. The error bars for the force dissymmetry, $g_z$, were calculated using the error propagation for uncorrelated variables (since $v_L$ and $v_R$ were measured independently). Namely, the standard deviation for $g_z$ was defined as
\[
\sigma_g = \sqrt{\left(\frac{\partial g_z}{\partial v_L}\sigma_L\right)^2+\left(\frac{\partial g_z}{\partial v_R}\sigma_R\right)^2} = \frac{4}{(v_L+v_R)^2}\sqrt{(v_R \sigma_L)^2+(v_L \sigma_R)^2}\,.
\]

\bibliographystyle{Science}
\bibliography{chiral}

\end{document}